+
\documentclass[aps,pra,twocolumn,floats,amsmath,amssymb,superscriptaddress]{revtex4-1}

\usepackage{graphicx}
\usepackage{dcolumn}
\usepackage{bm}

\usepackage{siunitx}
\usepackage{physics}
\usepackage{dsfont}
\usepackage{color}
\usepackage{notes2bib}

\begin{document}


\title{Fingerprints of stacking order in graphene layers \\ from \textit{ab initio} second-order Raman spectra}

\author{Albin Hertrich}
\email{albin.hertrich@physik.hu-berlin.de}
\author{Caterina Cocchi}%
\email{caterina.cocchi@physik.hu-berlin.de}
\author{Pasquale Pavone}%
\email{pasquale.pavone@physik.hu-berlin.de}
\author{Claudia Draxl}%
\email{claudia.draxl@physik.hu-berlin.de}
\affiliation{Institut f\"{u}r Physik and IRIS Adlershof, Humboldt-Universit\"{a}t zu Berlin, Berlin, Germany}
\affiliation{European Theoretical Spectroscopic Facility (ETSF)}

\date{\today}

\begin{abstract}
We present an \textit{ab initio} study based on density-functional theory of first- and second-order Raman spectra of graphene-based materials with different stacking arrangements and numbers of layers. 
Going from monolayer and bilayer graphene to periodic graphitic structures, we investigate the behavior of the first-order G-band and of the second-order 2D-band excited by the same set of photon energies. 
The former turns out to be very similar in all considered graphene-based materials, while in the latter we find the signatures of individual structures.
With a systematic analysis of the second-order Raman spectra at varying frenquency of the incident radiation, we monitor the Raman signal and identify the contributions from different phonon modes that are characteristic of each specific arrangement.
Supported by good agreement with experimental findings and with previous theoretical studies based on alternative approaches, our results propose an effective tool to probe and analyze the fingerprints of graphene-based and other low-dimensional materials.
\end{abstract}

\maketitle

\section{\label{sec:intro} Introduction}

Raman spectroscopy is one of the most used tools available for characterizing graphene-based materials. As an optical probe that gives access to both optical and vibrational properties, various information can be extracted from Raman experiments in a non-destructive way.
Since the first exfoliation of graphene sheets~\cite{graphene1}, Raman spectroscopy has been successfully used on graphene-based materials to determine 
information about the number of carbon layers and their stacking order~\cite{graf2007spatially,malard2009raman,hao2010probing,ferrari_layers,graphene_layer_out_of_plane,herz-maul13pssb},
as well as the presence of disorder~\cite{disorder,ferreira2010evolution}, 
strain~\cite{ni2008uniaxial,mohiuddin2009uniaxial,huang2009phonon,yoon2011strain,frank2011raman,neumann2015raman}, 
doping level~\cite{casiraghi2009doping,doping1, doping2,beams2015raman}, 
and much more~\cite{canccado2011quantifying,eckmann2012probing,ferrari_nano}. 

The state-of-the-art first-principles methodology for describing resonant Raman spectra from single and multiple phonon scattering of any order has been introduced more than two decades ago~\cite{proceedings,ambr+97zpb}. 
There, all required quantities are adopted from density-functional theory calculations and beyond.
Until now this approach has been adopted for computing first-order Raman spectra of many materials, including different kinds of superconductors~\cite{raman_draxl,pusc+01prb,ravi+03prb}, ladder compounds~\cite{spit+07prb,spit+08prb,spit+09njp},  and semiconductors also including excitonic effects~\cite{gillet_first, gillet_second, gillet_thesis}.
Later, an alternative formalism based on Feynman-diagrammatic approach has been proposed to obtain second-order Raman-scattering intensities from Fermi's golden rule~\cite{venezuela2011}.
This method is equivalent to the one presented in Ref.~\cite{proceedings} under certain conditions, namely that the dielectric function is calculated at the independent-particle level and that phonon energies are small~\cite{gillet_thesis}.
The Feynman diagrammatic approach has been used in the context of pseudopotential methods, to compute G~\cite{reic-wirt17prb} and 2D Raman bands of monolayer graphene~\cite{venezuela2011}, bilayer graphene~\cite{herziger,herz+18car}, and complex multilayer stacks including trilayer structures with different layer sequences~\cite{double-resonant-abc}.
Besides, Raman signatures of these systems have also been calculated with other theoretical approaches based on model Hamiltonians~\cite{thom-reic01prl,naru-reic08,naru-reic14,basko_2008,popo-lamb13prb,popov_carbon, popov_strain,popov_conference,kuku+14pssb,heller}.

Raman scattering intensities reflect the sensitivity of the polarizability of a material to the energy of the incoming light.
Since various experiments (see, \textit{e.g.}, Refs.~\cite{ferrari_layers,graf2007spatially,hao2010probing,herziger}) have been carried out using different excitation energies, it is often hard to distinguish whether the differences in the experimental observations result from this fact or from structural features, as the number of layers and the stacking.
Therefore, in contrast to previous theoretical works that have been focused on reproducing a specific experimental setup, we pursue here a different approach.
Adopting the method proposed in Ref.~\cite{proceedings} we study the fingerprints of first- and second-order Raman spectra of different graphene-based materials for the same set of excitation energies.
We provide a systematic analysis and comparison of the main features of representative graphitic structures including monolayer and bilayer graphene as well as graphite. 
Both the AA and AB stackings are considered.
In particular, we address the questions, how the stacking of graphene layers affects the main features in the Raman spectra, and how the spectra change at increasing excitation energies. 
We compute electronic structure, phonon bands, and dielectric tensors in the all-electron framework of density-functional theory (DFT) given by the \texttt{exciting} code~\cite{exciting}, which provides an efficient implementation of first- and second-order Raman scattering~\cite{albin-thesis} based on the theory in Ref.~\cite{proceedings}.
We focus on the behavior of the first-order G-band and of the second-order 2D-band, considering for the latter the dependence on the energy of the excitation beam.
We show and discuss how different phonon modes contribute to the overall spectral shape in different structures as a function of the excitation energy.

\section{\label{sec:method} Methodology}

\subsection{Raman-scattering efficiency}

The Raman-scattering efficiency for the transition between the initial $\left|i\right>$ and the final $\left|f\right>$ many-body states is connected to the transition matrix elements of the macroscopic dielectric tensor $\boldsymbol\varepsilon$ by the equation~\cite{proceedings,raman_draxl}:
\begin{equation}
  \label{eq:8}
  \boldsymbol{S}_{\omega_s}=\frac{N_\text{cell}\,V_{\!E}\,\omega_L\,\omega_s^3}{(4\pi)^2\,c^4}
\left|
      \mel**{f}{\boldsymbol\varepsilon_{\omega_L}}{i}
    \right|^2.  
\end{equation}
Here, $N_\text{cell}V_{\!E}$ is the phonon coherence volume, $c$ the speed of light, $\omega_L$ the excitation frequency, and $\omega_s$ the scattered light frequency with $\hbar\omega_s=\hbar\omega_L-\epsilon_f+\epsilon_i$.
The connection between the Raman-scattering efficiency and phonons can be expressed by introducing the phonon normal coordinates $\{Q_\mathbf{q}^j\}$, which quantify the atomic displacements associated with a phonon eigenmode $j$ at wave-vector $\mathbf{q}$.
The actual position of the atom $\alpha$ in the $l$th unit cell, $\mathbf{X}_l^\alpha$, can be written as the sum of a lattice vector $\mathbf{R}_l$, the equilibrium position  $\boldsymbol{\tau}^{\alpha}$, and the displacement out of equilibrium $\mathbf{u}_l^{\alpha}$.
The displacement corresponding to a phonon eigenmode is expressed as:
\begin{equation}
  \mathbf{u}_{l}^{\alpha }\!(Q_\mathbf{q}^j)=\frac{1}{\sqrt{N_\text{cell}\,M_\alpha}}\,\mathbf{w}^{j\alpha}_{\mathbf{q}}\,
  e^{\mathrm{i}\mathbf{q}\cdot \mathbf{R}_l}\,Q^j_{\mathbf{q}}~.
  \label{eq:37}
\end{equation}
In Eq.\:\eqref{eq:37}, $M_\alpha$ is the mass of atom $\alpha$ and $\mathbf{w}_\mathbf{q}^\alpha$ the normalized eigenvector of the dynamical matrix
  \begin{equation}
    D^{\gamma \gamma'}_\mathbf{q}=\frac{1}{N_\text{cell}\sqrt{M_\gamma\, M_{\gamma'}}}\sum_{ll'}
  \left. 
    \frac{\partial^2 E_\text{tot}}{\partial u_l^\gamma\, \partial u_{l'}^{\gamma'}}
  \right|_{u=0} 
    \, e^{\mathrm{i}\mathbf{q}\cdot (\mathbf{R}_{l'}-\mathbf{R}_{l})}~,
\label{eq:dyn}
\end{equation}
with the combined Cartesian-atomic index  $\gamma$.
Using the Taylor expansion of the dielectric function
\begin{align}
  \label{eq:4}
  \nonumber
  \boldsymbol\varepsilon_{\omega_L}\left(Q\right)=&\left . \boldsymbol\varepsilon_{\omega_L}\right|_{Q=0}
  +\sum_{{j \mathbf{q}}}\left . \frac{\partial \boldsymbol\varepsilon_{\omega_L}}
  {\partial Q_{\mathbf{q}}^j} \right|_{Q=0} \! Q_{\mathbf{q}}^j \\\nonumber
  &+\frac{1}{2}\sum_{j j' \mathbf{q} \mathbf{q}'}\left .  \frac{\partial^2\boldsymbol\varepsilon_{\omega_L}}
  {\partial Q_{\mathbf{q}}^j\,\partial Q_{\mathbf{q}'}^{j'}} \right|_{Q=0} \! Q_{\mathbf{q}}^j\, Q_{\mathbf{q}'}^{j'} \\[1.5mm]
  &+\mathcal{O}(Q^3)~,
\end{align}
the matrix elements in Eq.\:(\ref{eq:8}) separate into terms of different order in the phonon normal coordinates:
\begin{equation}
  \label{eq:41}
      \mel**{f}{\boldsymbol\varepsilon_{\omega_L}}{i}
  =  \left[
    \boldsymbol\varepsilon_{i\rightarrow f}
  \right]^{{\kern -0.3mm}(1)}\!\!(Q)
  +  \left[
    \boldsymbol\varepsilon_{i\rightarrow f}
  \right]^{{\kern -0.3mm}(2)}\!\!(Q^2)
  +\mathcal{O}(Q^3)~.
\end{equation}
Here, $Q$ stands for the set of all $Q_\mathbf{q}^j$.
Within the adiabatic approximation, the initial and final state are products of the electronic and vibrational states. 
The derivatives of the dielectric function at zero displacement depend only on the electronic states, while the phonon normal coordinates depend only on the vibrational states, labeled $\ket{\mu}$ and $\ket{\nu}$ in the following.
Before the absorption and after the emission of the photon, the system is in the electronic ground state, and the transition takes place in the vibrational state only. 
In the optical limit, the momentum of the photon is negligible. Because of momentum conservation, the sum of all created phonons has to be zero.
Under these conditions, the first- and second-order contributions in Eq.\:(\ref{eq:41}) are given~by
\begin{align}
  \label{eq:29a}
  \left[
    \boldsymbol\varepsilon_{i\rightarrow f}
  \right]^{(1)}
= & \left[
  \boldsymbol\varepsilon_{\mu\rightarrow\nu}
  \right]^{(1)}
  = \sum_j \left. \frac
      {\partial\boldsymbol\varepsilon_{\omega_L}}
      {\partial Q_0^j}\right|_{Q=0}
      \mel**{\nu}{Q_0^j}{\mu}
  \\ \nonumber
  \left[
    \boldsymbol\varepsilon_{i\rightarrow f}
  \right]^{(2)}
= & \left[
  \boldsymbol\varepsilon_{\mu\rightarrow\nu}
  \right]^{(2)}\\
  = & \frac{1}{2}\sum_{jj'\mathbf{q}} \left. \frac
      {\partial^2\boldsymbol\varepsilon_{\omega_L}}
      {\partial Q_\mathbf{q}^j \,\partial Q_{-\mathbf{q}}^{j'}}\right|_{Q=0}
      \mel**{\nu}{Q_\mathbf{q}^j \,Q_{-\mathbf{q}}^{j'}}{\mu}.
      \label{eq:29b}
\end{align}
The two ingredients in Eqs.\:(\ref{eq:29a}) and (\ref{eq:29b}), namely the derivatives of the dielectric function with respect to the phonon normal coordinates and the phonon matrix elements, can be calculated individually. 
Due to momentum conservation, in first-order Raman processes only $\Gamma$-point phonons contribute, while the second-order spectrum is generally made up by contributions of any two phonons with opposite momenta within the first Brillouin zone (BZ). 
For the Raman-scattering efficiency, a summation over all final vibrational states has to be calculated. Including thermal occupation of the initial vibrational state, the full Raman-scattering intensity is given~by
\begin{equation}
  \label{eq:42}
  \boldsymbol{S}_{\omega_s}=\frac{\displaystyle N_{\rm cell}\,V_{\!E}\,\omega_L\,\omega_s^3    \sum_{\mu} e^{-E_\mu/k_BT}
    \sum_{\nu }
    \left| \sum_n
      \left[ \boldsymbol\varepsilon_{\mu\rightarrow \nu} \right]^{(n)}
    \right|
    ^2}
  {\displaystyle(4\pi)^2\,c^4  \sum_{\mu} e^{-E_\mu/k_BT}}~.
\end{equation}
This expression corresponds to the integrated intensity of the Raman line, which is then broadened with a Lorentzian function.

\subsection{Derivatives of dielectric function}
The derivatives of the macroscopic dielectric function with respect to the phonon normal coordinates in Eqs.\:(\ref{eq:29a}) and (\ref{eq:29b}) are calculated in the \textit{frozen-phonon approximation}, where the displacements of the atoms are treated as a quasi-static perturbation.
Instead of the complex eigendisplacements given in Eq.\:(\ref{eq:37}), the real combination corresponding to phonons at $\mathbf{q}$ and $\mathbf{-q}$ 
\begin{align}
  \label{eq:32}
  \nonumber
  \mathbf{u}_{l}^{\alpha }\!( \bar{Q}_\mathbf{q}^j)=&\frac{1}{2\sqrt{N_\text{cell} \,M_{\alpha}}}
  \left(
    \mathbf{w}^{j\alpha}_{\mathbf{q}}\,e^{\mathrm{i}\mathbf{q}\cdot\mathbf{R}_l}+
    \mathbf{w}^{j\alpha}_{-\mathbf{q}}\,e^{-\mathrm{i}\mathbf{q}\cdot\mathbf{R}_l}
  \right)
  \bar{Q}_\mathbf{q}^j \\
  =&\frac{1}{\sqrt{N_\text{cell}\, M_{\alpha}}}
  \,\real \!  \left(
    \mathbf{w}^{j\alpha}_{\mathbf{q}}\,e^{\mathrm{i}\mathbf{q}\cdot\mathbf{R}_l}
  \right)
  \bar{Q}_\mathbf{q}^j
\end{align}
is used with the real normal coordinate $\bar{Q}_\mathbf{q}^j$.
The dielectric function is calculated for several distorted geometries and a polynomial fit is applied:
\begin{equation}
  \label{eq:7}
  \boldsymbol\varepsilon= \boldsymbol\varepsilon^{(0)}+{\boldsymbol\varepsilon_\mathbf{q}^{j\,(1)}} \,\bar{Q}_\mathbf{q}^j+{\boldsymbol\varepsilon_\mathbf{q}^{j\,(2)}} \,({{\bar{Q}_\mathbf{q}}^j)^2 }  +...~ .
\end{equation}
For $\Gamma$-point phonons the real and complex displacements in Eqs.~\eqref{eq:32} and~\eqref{eq:37}, respectively, are equal and the first derivative of the dielectric function is given by the first-order coefficient
\begin{equation}
\left.  \frac{\partial\boldsymbol\varepsilon}{\partial Q_0^j}\right|_{Q=0}\!\!=\,{\boldsymbol\varepsilon_0^{j\,(1)}}\,.
  \label{eq:43}
\end{equation}
To compute second-order Raman spectra in general, displacements along different phonon modes have to be combined. 
With the restriction to overtone spectra, in the absence of contributions from different phonon bands in one process, the one-dimensional fit of the dielectric function given by Eq.\:(\ref{eq:7}) is sufficient. 
In this case, the second derivative with respect to the normal coordinates of phonons with momentum $\mathbf{q}$ and $-\mathbf{q}$ of the same phonon band $j$ is given by the second-order coefficient
\begin{equation}
  \label{eq:43a}
 \left. \frac{\partial^2\boldsymbol\varepsilon}{\partial Q_\mathbf{q}^j \, \partial Q_\mathbf{-q}^j}\right|_{Q=0}\!\!=\,4\, 
  {\boldsymbol\varepsilon_\mathbf{q}^{j\,(2)}}~.
\end{equation}

\subsection{Vibrational matrix elements}
 The vibrational matrix elements defined in Eqs.\:(\ref{eq:29a}) and (\ref{eq:29b}) can be evaluated numerically by using a polynomial fit for the total energy (or the scalar product of forces and eigenvectors), in analogy to Eq.\:(\ref{eq:7}).
 In crystalline systems, due to the large coherence volume of the phonons, the main contributions come from the harmonic terms. Neglecting all higher-order terms, the vibrational matrix elements defined in Eqs.\:(\ref{eq:29a}) and (\ref{eq:29b}) can be calculated analytically. For simplicity, let us consider for the moment only one phonon mode, with frequency $\omega$ and occupation number $\nu$ in the vibrational state $\ket{\nu}$.
The phonon normal coordinate, written in terms of creation- and annihilation-operators ($\hat{\bm{a}}^\dagger\!$ and $\hat{\bm{a}}$), is given by
\begin{equation}
  \label{eq:44}
  Q =\frac{1}{\sqrt{2\,\omega}}\,\left(\hat{\bm{a}}+\hat{\bm{a}}^\dagger\right)~.
\end{equation}
Inserting Eq.\:(\ref{eq:44}) into the first-order phonon matrix elements given in Eq.\:(\ref{eq:29a}), we obtain 
\begin{align}\nonumber
  \mel**{\nu}{Q}{\mu}=&\frac{1}{\sqrt{2\,\omega}}
                        \mel**{\nu}{\hat{\bm{a}}+\hat{\bm{a}}^\dagger}{\mu}\\
  =&\frac{1}{\sqrt{2\,\omega}}
     \left[
     \sqrt{\mu}\,\delta_{\nu,\mu-1}+\sqrt{\mu+1}\,\delta_{\nu,\mu+1}
     \right]~.
     \label{eq:45}
\end{align}
Here, the first term belongs to an anti-Stokes process, while the second one to a Stokes process.
Using the thermal occupation number
\begin{equation}
  \bar{n}_\mathbf{q}^j=\frac{1}{e^{-\hbar\omega_\mathbf{q}^j/k_BT}-1} ~ ,
\end{equation}
we obtain from Eq.\:(\ref{eq:42}) 
\begin{align} \nonumber
  {\boldsymbol{S}^{(1)}_{\omega_s}}=\frac{N_{\rm cell}\,V_{\!E}\,\omega_L\,\omega_s^3    }
  {(4\pi)^2\,c^4  } \sum_{j} \frac{1}{2\,\omega_0^j}
  \left|{\boldsymbol\varepsilon_0^{j\,(1)}}\right|^2
  \,(\bar{n}_0^j+1)
  \\ \times \, \delta(\omega_L-\omega_s-\omega_0^j) 
  \label{eq:12}
\end{align}
for the first-order Stokes Raman spectrum within the harmonic approximation.
Likewise, the harmonic second-order overtone spectrum is given by
\begin{align} \nonumber
  {\boldsymbol{S}^{(2)}_{\omega_s}}=\frac{N_{\rm cell}\,V_{\!E}\,\omega_L\,\omega_s^3    }
  {(4\pi)^2\,c^4  } \sum_{\mathbf{q}j} &\left(\!2\,\omega_\mathbf{q}^j\right)^{\!\!-2} 
                                         \left| 2\, {\boldsymbol\varepsilon_\mathbf{q}^{j\,(2)}} \right|^2 
  \, (\bar{n}_\mathbf{q}^j+1)^2
  \\ & \times \,
  \delta(\omega_L-\omega_s-2\,\omega_\mathbf{q}^j)  ~.
  \label{eq:19}
\end{align}
Here, the relation $\omega_\mathbf{q}^j=\omega_\mathbf{-q}^j$ is used.

\subsection{Brillouin zone interpolation \label{sec:gaussian}
}
For the full calculation of the second-order (overtone) Raman spectrum, an interpolation over the BZ is necessary. 
The straightforward way to do so is to calculate the second derivatives of the dielectric function on a coarse grid of $\mathbf{q}$-points and, for each excitation energy, use Fourier interpolation to obtain the derivatives on a denser grid. 
This method works well when the second derivatives vary slowly. 
This is not the case in graphene-based materials, where the second-derivatives with respect to the transverse optical phonon modes are resonant at certain excitation energies, depending on the phonon momenta. For this reason, this kind of interpolation does not give the correct resonance behavior. 

To overcome this problem, another interpolation technique, tailored to this use case, is adopted.
Following Eq.\eqref{eq:43a}, the square modulus of the second derivatives of the dielectric function is approximated by a sum of Gaussian functions centered at the resonance positions:
\begin{equation}
    \label{eq:g1}
    \left|4\,\boldsymbol\varepsilon_\mathbf{q}^{(2)}\!(\omega_L)\right|^2
      \approx \, \sum_{r} \,
    \frac{A_{r}}{\sigma_{\!r}\sqrt{2\,\pi}}\;e^{-\frac{(\omega_L-\omega_{r})^2}{2\,{\sigma_{\!r}}^2}}.
  \end{equation}
 The Gaussian parameters $A_{r}$, $\sigma_{\!r}$, and $\omega_{r}$ are obtained from a least squares fit.
 Instead of interpolating the derivatives for each excitation energy, the Gaussians parameters are interpolated on the \textbf{q-}point grid. Assuming that the parameters vary slowly, the main features of the derivatives of the dielectric function between the calculated points can be captured with high accuracy.

\section{Results}

\subsection{\label{graphene} Electronic, optical, and vibrational properties}
In this section, we present the structural, optical, and vibrational properties that are the necessary ingredients for accessing and analyzing the first- and second-order Raman spectra.
For a comprehensive study on the properties of graphene and related materials we refer the readers to, \textit{e.g.}, Refs.~\onlinecite{handbook1,bilayer_electronic_structure}.

\begin{figure}[hbt]
  \includegraphics{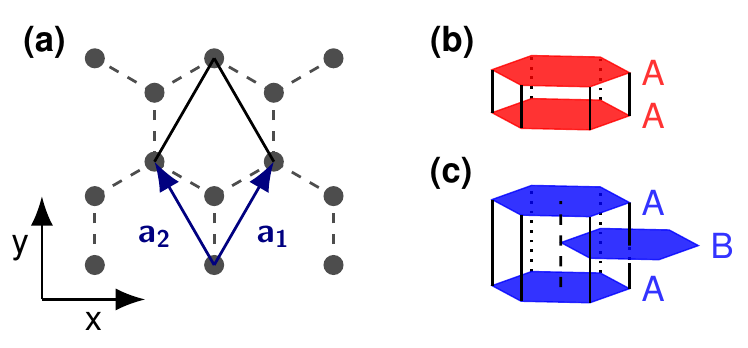}
  \caption{\label{fig:lattice} (Color online)  (a) Unit cell of graphene with basis vectors $\mathbf{a}_1$ and $\mathbf{a}_2$. (b) AA and (c) AB stacking of graphene sheets. }
\end{figure}

In the AA, or simple hexagonal stacking (Fig.\:\ref{fig:lattice}b) of graphene sheets (Fig.\:\ref{fig:lattice}a), each atom of a layer is situated directly on top of an atom of the neighboring layer.
AA graphite is considered here to investigate the AA stacking in the limit of many layers, even though this material itself is metastable~\cite{aa_stability,aa_stability_2}. 
In the AB, or Bernal stacking (Fig.\:\ref{fig:lattice}c), every second layer is shifted so that there is an atom on top of each carbon ring of the neighboring layers.
AB stacking is energetically favorable~\cite{bilayer_binding_qmc}. This is the most common configuration of monocrystalline graphite.

\begin{figure}[htb]
  \includegraphics{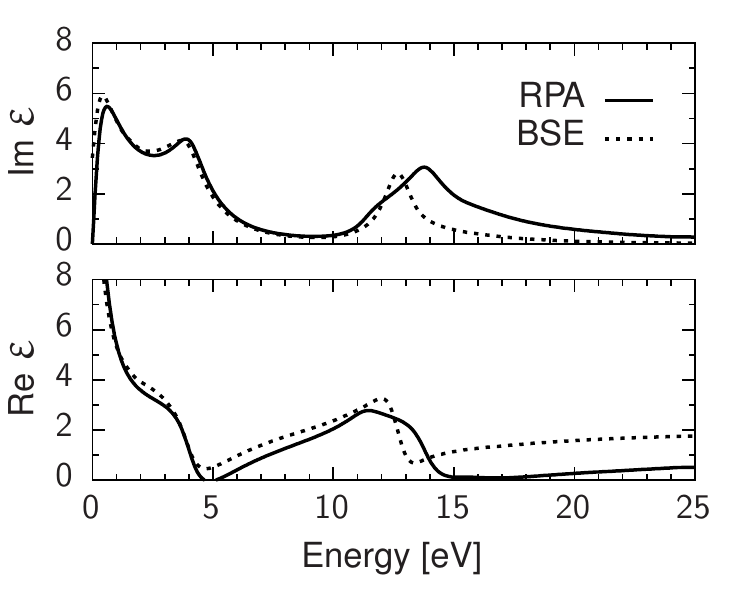}
  \caption{\label{fig:epsilon} 
Real (lower panel) and imaginary (upper panel) part of the in-plane component of the dielectric function of graphene calculated within the RPA and by solving the Bethe-Salpeter equation (BSE).}
\end{figure}

Graphene is a peculiar semi-metal characterized by a linear band dispersion in the vicinity of the K point~\cite{graphene_dirac}.
The AA stacking conserves the linear band dispersion, but vertically shifts the  bands~\cite{bilayer_electronic_structure}, while the AB stacking leads to parabolic $\pi$ and $\pi^*$ bands.
Due to the strong electronic screening coming from the $\pi$ and $\pi^*$ bands, the electron-hole interaction only plays a minor role in the dielectric response in graphene-based materials in the optical region ($1-\SI{4}{eV}$).
Hence, the dielectric function is usually well-described within the random-phase approximation (RPA)~\cite{bse_trevisanutto,graphene_bse,graphene_bse2}, as shown below in Fig.~\ref{fig:epsilon} in comparison with the result obtained from the solution of the Bethe-Salpeter equation (BSE)~\cite{bse,bse_strinati,bse_rohlfing}.
The formalism for calculating the dielectric function with the RPA is described in Appendix A. 
Since the out-of-plane component does not contribute to Raman scattering due to the symmetry of the Raman-active phonons, only the in-plane component is shown in Fig.\,\ref{fig:epsilon} and discussed below.
In the low-energy region ($E<\SI{5}{eV}$), where contributions mainly come from transitions between $\pi$ and $\pi^*$ bands, our results confirm that the main spectral features are well captured within the RPA.  
In the high-energy region, above \SI{5}{eV}, electron-hole correlation plays a more significant role, and the difference between the RPA and BSE results is enhanced. 
However, for Raman spectra in the visible and near UV region those spectral features have no influence, and therefore a description within the RPA is sufficient for the purpose of this paper.
Given the similarities of the electronic properties of all examined graphene-based materials, the results shown here for graphene are representative to those of the other materials considered in this work.

\begin{figure}[htb]
  \includegraphics{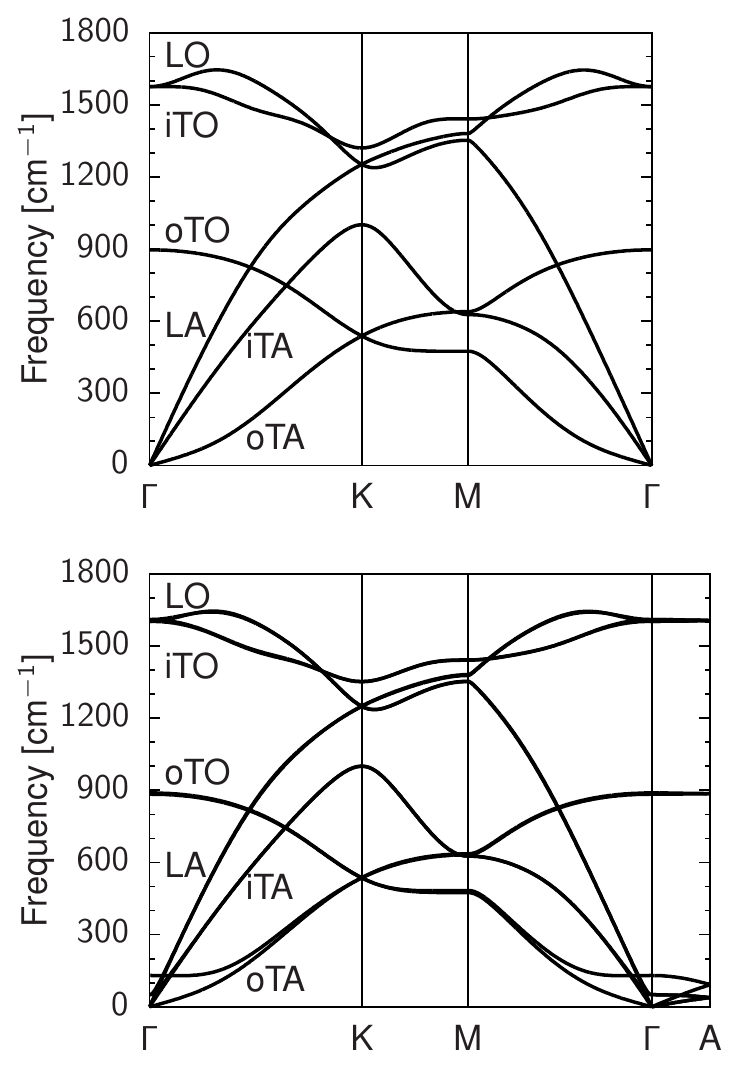}
  \caption{\label{fig:phonon} LDA phonon dispersion of graphene (upper panel) and AB graphite (lower panel).}
\end{figure}

The phonon dispersion of graphene (Fig.\:\ref{fig:phonon}, upper panel) exhibits in-plane and out-of-plane modes. 
Because of the presence of two atoms in the unit cell, there are six phonon bands belonging to longitudinal optical (LO), in-plane transverse optical (iTO), out-of-plane transverse optical (oTO), longitudinal acoustical (LA), in-plane transverse acoustical (iTA), and out-of-plane transverse acoustical (oTA) vibrations.
For both optical and acoustical phonons in the whole BZ, frequencies of the in-plane modes are higher than the corresponding out-of-plane ones. 
DFT phonon frequencies calculated in the local-density approximation (LDA) are in overall good agreement with experimental values from inelastic x-ray scattering~\cite{phonon_exp1,phonon_exp2,phonon_exp3}, except for iTO phonons near the K-point, which are overestimated by approximately $4\%$. 
This behavior has been ascribed to the underestimation of the electron-phonon coupling for iTO phonons at K by DFT with local or semilocal exchange-correlation functionals~\cite{kohn0}.
While in the direct vicinity of K an improved description is obtained from \textit{GW}~\cite{phonon_gw}, a correction that is valid for the full BZ has been realized so far only by means of empirical parameters~\cite{venezuela2011, herziger}. 
As the error in the phonon dispersion does not affect the essence of our results, we use the LDA frequencies.

Stacking of graphene layers mainly preserves the vibrational properties. 
The phonon dispersion of AA graphite is almost identical to the one of graphene with slightly higher iTO frequencies around K.
Since both bilayer graphene and AB graphite have four atoms in the unit cell, there are overall 12 phonon bands. 
Due to the weak inter-layer interaction, every phonon mode of graphene is split up into two phonon modes belonging to in-phase and out-of-phase vibrations of the sub-lattices. The frequencies of these modes are nearly degenerate, except for the oTA modes near the $\Gamma$ point, where the splitting is much larger.
For AA bilayer graphene the splitting is up to \SI{3}{cm^{-1}}, for AB stacking up to \SI{10}{cm^{-1}}.

In graphene and AA graphite, the degenerate in-plane optical $\Gamma$-point phonons have $E_{2g}$ symmetry and are thus first-order Raman active.
In bilayer graphene and AB graphite, phonon modes are split into in-phase and out-of-phase vibrations of the layers. 
This leads to four Raman-active phonon modes, the high-energy in-phase and the low-energy out-of-phase modes. 
In the present work, the high-energy modes are investigated.
As shown below, major contributions to the second-order spectra come from iTO phonons along $\Gamma$-K.

\subsection{\label{sec:1st}First-order Raman spectra: The G-band}

\begin{figure}[hbt]
  \includegraphics{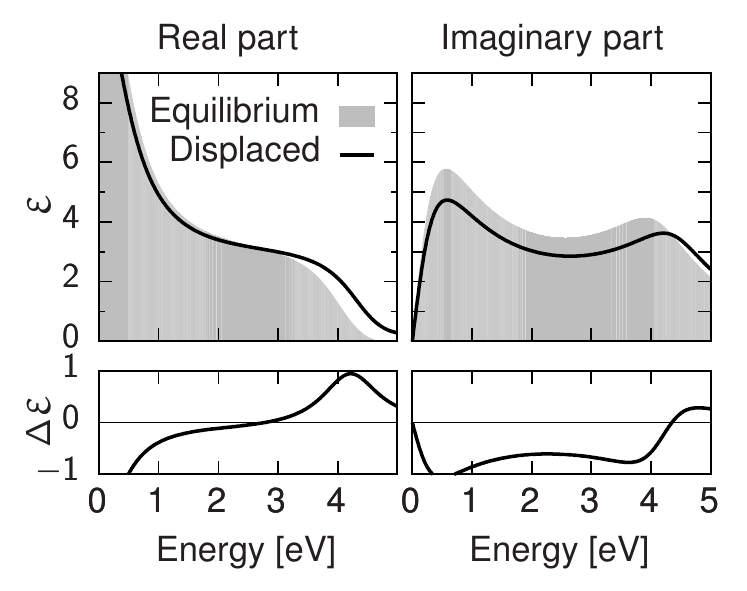}
  \caption{\label{fig:epsilon_first} Upper panels: Diagonal in-plane component of the real and imaginary part of the dielectric function of graphene. Shaded areas correspond to the equilibrium geometry; solid lines correspond to a structure with atoms displaced along the $y$ direction. Lower panels: Difference between the equilibrium and displaced dielectric functions shown in the upper panels.}
\end{figure}

First-order Raman spectra always show discrete peaks around the Raman-active $\Gamma$-point phonon frequencies. The G-band comes from a pair of degenerate optical phonons, leading to one single peak.
Due to degeneracy, any opposite in-plane displacements of the two atoms in the unit cell corresponds to a phonon eigendisplacement.
As shown in Fig.\:\ref{fig:epsilon_first}, the dielectric tensor is affected by the displacements in the whole optical region. 
While a relative displacement along the $x$ direction (according to the reference axes in Fig.\ref{fig:lattice}a) changes the off-diagonal components, a relative displacement along the $y$ direction (see Fig.\ref{fig:lattice}a) changes the diagonal components. 
For this reason, the outscattered light is unpolarized, regardless of the polarization of the incoming light.

\begin{figure}[h!bt]
  \includegraphics{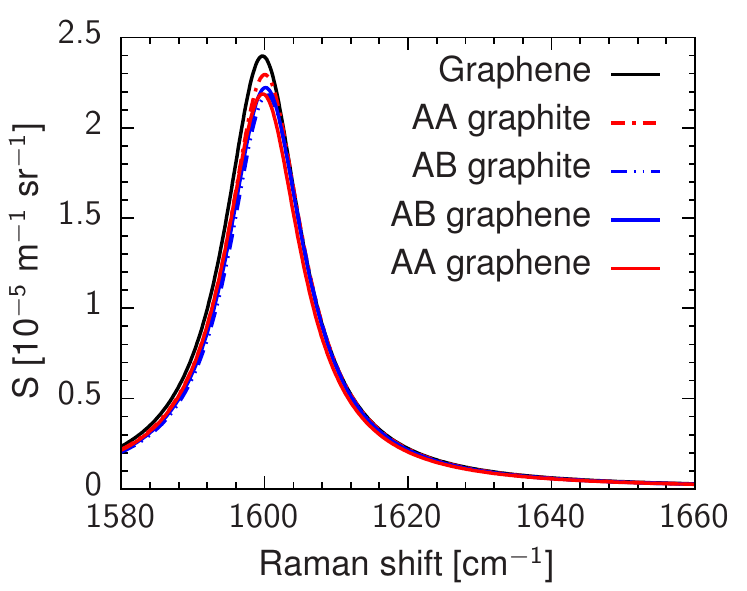}
  \caption{\label{fig:first} (Color online)  First-order Raman spectra of graphene as well as of graphite and bilayer graphene in the AA and AB stacking sequences. }
\end{figure}

The Raman-scattering efficiencies for graphene, bilayer graphene, and graphite in AA and AB stacking are calculated from the first derivatives of the dielectric function and the phonon frequencies (Eq.\:\ref{eq:12}).
The Raman spectra shown in Fig.\:\ref{fig:first} are obtained by applying a Lorentzian broadening of \SI{13}{cm^{-1}} that corresponds to the experimentally measured linewith~\cite{g_width}. 
By comparing the first-order G-band in the spectra of the investigated graphene-based materials (Fig.\:\ref{fig:first}), we see only very small shifts in the resonance position, coming from variations in the $\Gamma$-point phonon frequencies. 
For an efficient characterization of graphene-based materials, these differences are not sufficient and the second-order Raman spectra need to be considered, as already demonstrated experimentally (see, \textit{e.g.}, Ref.~\cite{graf2007spatially}).

\begin{figure}[h]
  \centering
  \includegraphics{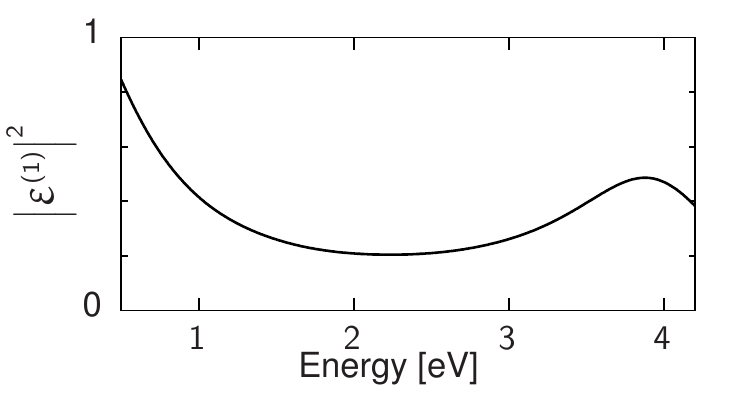}  
  \caption{Resonance behavior of the first-order Raman spectra of graphene.}
  \label{fig:first_res}
\end{figure}

In Fig.\:\ref{fig:first_res} we show the energy dependence of the first derivative of the dielectric function of graphene.
While the intensity is strongly enhanced in both the IR (below $\SI{1}{eV}$) and in the UV region (at about $\SI{3.8}{eV}$), in the visible range it is almost flat.
Hence, at visible frequencies the Raman scattering intensity is dominated by the energy prefactor $\omega_L\,\omega_s^3$ of Eq.\:(\ref{eq:12}).
This behavior is consistent with the measured Raman differential cross section of graphene in the visible region, which indeed exhibits a dependence on the fourth power with respect to the energy~\cite{canc+07prb}.

\subsection{\label{sec:2nd} Second-order Raman spectra: The 2D-band}

\begin{figure}[h!]
 \includegraphics{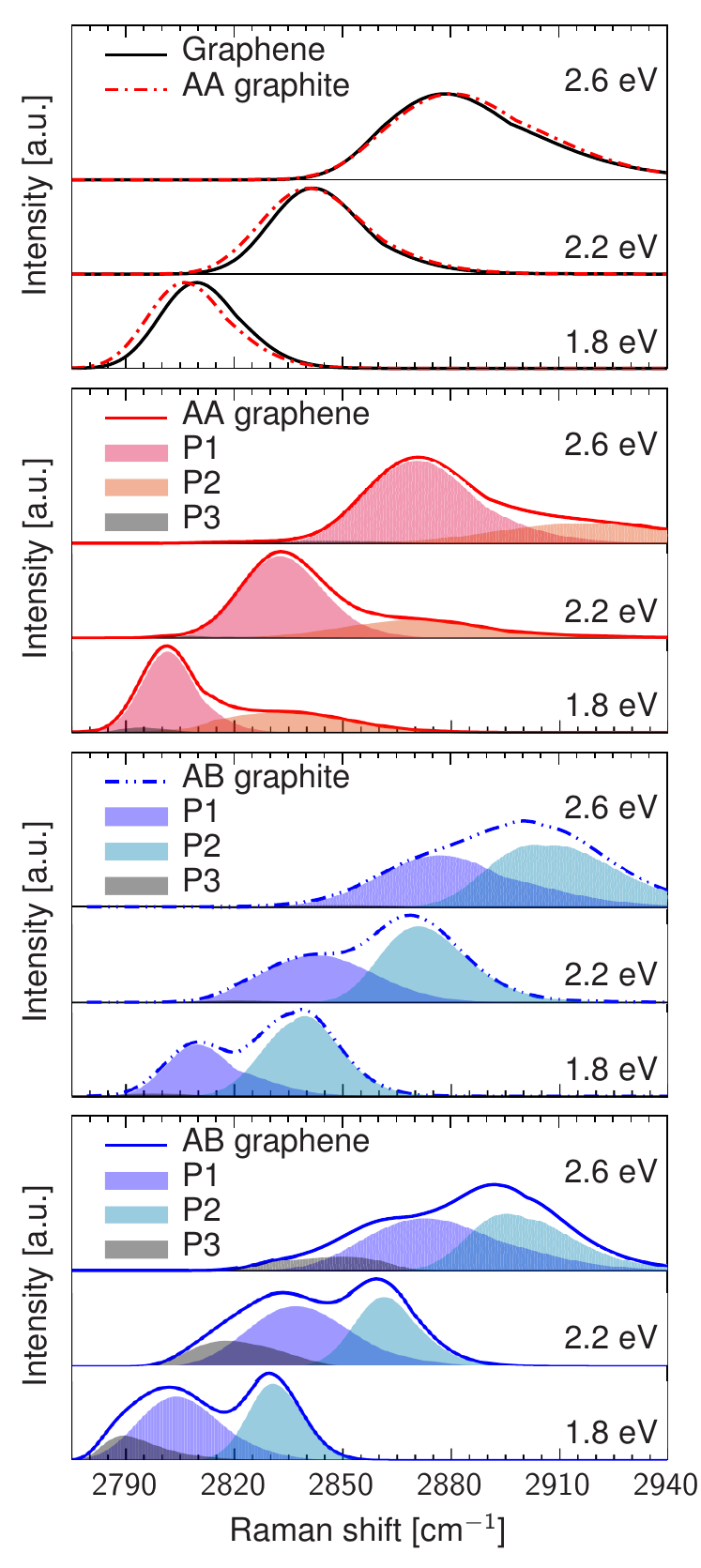}
 \caption{\label{fig:second}(Color online) Normalized second-order Raman 2D band of graphene, AA graphite and bilayer graphene, and AB graphite, and bilayer graphene for excitation energies \SI{1.8}{eV}, \SI{2.2}{eV}, and \SI{2.6}{eV}. The filled areas under the curves correspond to the contributions of the three resonances in Fig.~\ref{fig:second1}, lower panels.}
\end{figure}

In graphene-based materials, second-order Raman spectra are dominated by few resonant contributions from overtones of the iTO phonons.
LO modes, on the contrary, do not contribute.
The 2D-band, built up by these resonant vibrational contributions is presented in Fig.~\ref{fig:second} for the case of graphene, bilayer graphene (AA and AB stacking), and graphite (AA and AB stacking). 
For a consistent comparison, excitations corresponding to the same photon frequencies have been considered in all cases.
The spectra are calculated by applying Eq.\:\eqref{eq:19} with the Gaussian interpolation technique described above for phonons along the $\Gamma$-K line. 
In the following, we use the notation $\mathbf{q}=\lambda \text{K}$ with $\lambda \in [0,1]$ for points in reciprocal space on this line.
  
  \begin{figure}[h!]
  \includegraphics{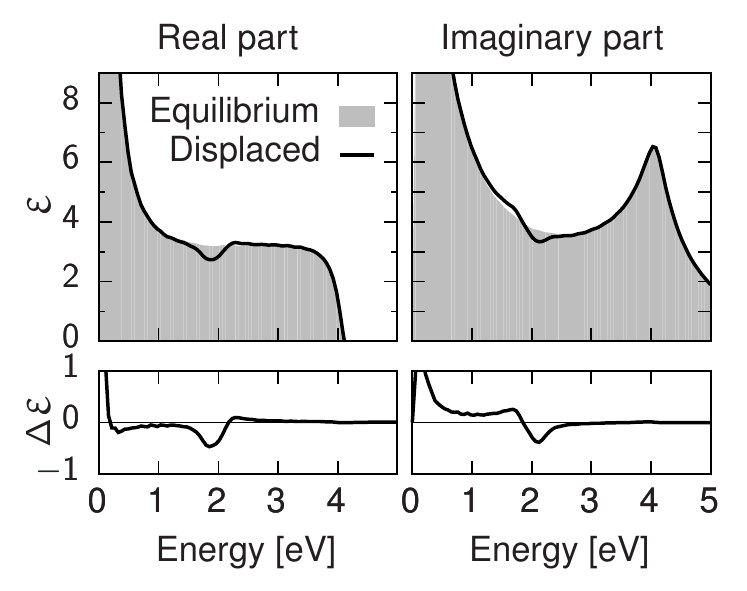}
  \caption{\label{fig:eps_second}(Color online) Upper panels: Diagonal in-plane component of the real and imaginary part of the dielectric function of graphene. Shaded areas correspond to the equilibrium geometry; solid lines correspond to a structure with atoms displaced along the iTO phonon at $\mathbf{q}=\SI{3/4}{K}$. Lower panels: Difference between the equilibrium and displaced dielectric functions shown in the upper panels.}
\end{figure}
  
\begin{figure}[hbt]
    \includegraphics{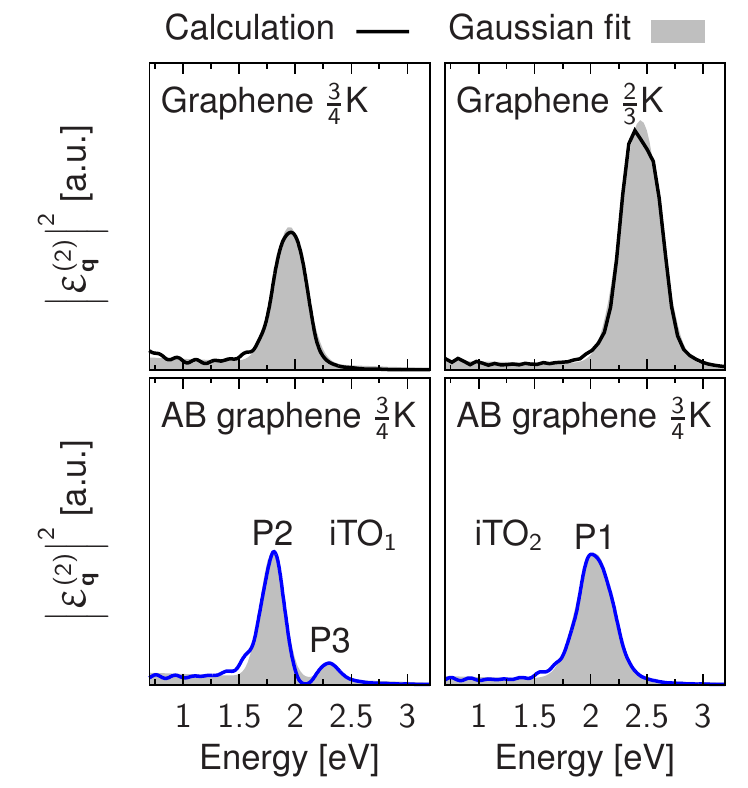}
    \caption{\label{fig:second1} Square modulus of the second derivatives of the dielectric function with respect to the iTO phonons normal coordinates for graphene at $\mathbf{q}=\SI{3/4}{K}$ (top left), graphene at $\mathbf{q}=\SI{2/3}{K}$ (top right), as well as for the lower (bottom left) and the higher (bottom right) iTO mode of AB bilayer graphene at  $\mathbf{q}=\SI{3/4}{K}$. Solid lines correspond to calculated values while shaded areas to the Gaussian fits.
}
  \end{figure}

First of all, we notice that the 2D-band is dispersive, since different phonon modes are in resonance at different excitation energies. As shown in Fig.~\ref{fig:eps_second}, displacements along the iTO modes cause significant changes in the dielectric function in a small energy region. These changes lead to resonances in the square modulus of the second derivatives of the dielectric function (see Fig.~\ref{fig:second1}).
Thus, peaks in the second-order Raman spectra are at twice the frequency of the corresponding phonon modes.
As both the positions of these resonances (upper panels of Fig.~\ref{fig:second1}), and the corresponding phonon frequencies (Fig.~\ref{fig:phonon}) decrease when approaching the K-point of the BZ, the position of the 2D-band maximum increases with the excitation energy by 80-\SI{90}{cm^{-1}/eV}, in excellent agreement with the experimental one of \SI{88}{cm^{-1}/eV} from Ref.~\cite{ferrari_layers}, \bibnote{While absolute values of phonon frequencies are systematically overestimated by LDA compared to experiments, such (relatively small) errors largely compensate each other when computing differences.}.

 The 2D-band of graphene and AA graphite (Fig.~\ref{fig:second}, top panel) is formed by a single peak, since in these structures there is only one resonance corresponding to each iTO phonon mode. 
 The spectra of these two systems almost overlap. 
 The slight shift by a few cm$^{-1}$ appearing at the excitation energy of 1.8 eV almost vanishes at increasing photon frequencies.
  The 2D-bands of the other materials have a fine-structure consisting of three sub-peaks, labeled P1, P2, and P3, which contribute differently for varying stacking sequences and excitation energies. 
  P1 comes from a resonance in the higher iTO phonon band, while P2 and P3 stem from resonances in the lower one (see lower panels of Fig.~\ref{fig:second1}).
  In terms of Raman shifts, the 2D bands of all stackings follow the same trend, with the resonance moving to higher wave numbers at increasing photon frequency.
  Likewise, peaks become broader upon larger photon energies.
  The most remarkable difference between the considered layer arrangements is given by the weight of the three sub-peaks.
  Among all, P3 gives the smallest contribution to the 2D-band.
  In the spectra corresponding to the excitation energy $\hbar \omega_L=\SI{1.8}{eV}$ of AA bilayer graphene and AB graphite, it only provides a very small contribution that is hardly visible in Fig.~\ref{fig:second}.
  Conversely, in AB graphene (Fig.~\ref{fig:second}, bottom panel) the weight of P3 is most prominent, especially at lower photon frequencies ($\hbar \omega_L=\SI{1.8}{eV}$ and $\hbar \omega_L=\SI{2.2}{eV}$), where its contribution is of the same order as the other sub-peaks.
  
  Another relevant difference between the second-order Raman spectra of these stackings is the relative weight of P1 and P2.
  In AA bilayer graphene, P1 dominates the spectral shape, while P2 appears weak and broad for all considered photon frequencies.
  As shown in Fig.~\ref{fig:second1}, this behavior indicates the predominant contribution of the iTO$_2$ mode to the Raman scattering process in this structure.
 Such one-peak structure of the 2D-band in AA bilayer graphene resembles the resonant shape exhibited by the AB bilayer stacking at very large twist angles~\cite{kim+12prl,have+12nl,jori-canc13ssc}.
 In turn, it is also very similar to the spectral shape of monolayer graphene, shown in the top panel of Fig.~\ref{fig:second}.
 Interestingly, in the twisted bilayer structure, the intensity of the 2D-band significantly increases at lower angles due to the increasing overlap between the Dirac cones~\cite{kim+12prl,have+12nl,jori-canc13ssc}.
 This feature cannot be appreciated in our plots, due to the missing prefactor in the calculation of resonance intensities.
  On the other hand, in the AB stacking of both bilayer graphene and graphite, P2 is more intense than P1 especially at $\hbar \omega_L=\SI{1.8}{eV}$ and $\hbar \omega_L=\SI{2.2}{eV}$.
  For the largest considered photon energy ($\hbar \omega_L=\SI{2.6}{eV}$) the height of the two peaks is substantially equivalent in both structures.
 All second-order Raman spectra displayed in Fig.~\ref{fig:second} exhibit analogous trends in terms of shifts of the 2D-band.
  Their sensitivity to the stacking is significantly enhanced compared to the first-order G-band (see Fig.~\ref{fig:first_res}).
Specific resonance features in graphene, AA bilayer graphene, AB bilayer graphene and AB graphite are clearly distinguishable. 
In the limit of infinite stacking in the AA sequence, the 2D-band closely resembles the one of graphene, as can be seen in the similarity of the spectra of AA graphite and graphene.

The trends shown here are in line with earlier experimental observations and support their interpretation~\cite{ferrari_layers,disorder,malard2009raman,zhan+16ns,herziger}.
The one- and two-peak spectral shape of the 2D Raman band of graphene and AB graphite, respectively, is in agreement with the results of Ferrari and coworkers~\cite{ferrari_layers,disorder}.
Also the blue-shift of the 2D-band upon increasing photon frenquency is well known, as shown for AB graphite~\cite{zhan+16ns} and AB bilayer graphene~\cite{herziger}.
Quantitatively, the computed Raman shifts presented in this work are systematically slightly blue-shifted compared to the measured counterparts, due to adoption of the LDA for the calculation of both electronic structure and phonon frequencies.
For the same reason, our calculated Raman shifts are somewhat larger than those reported in previous first-principles studies on the same materials~\cite{venezuela2011,herziger,popov_carbon,double-resonant-abc,popov_conference} where the $GW$ approximation was adopted.
Also, in the works cited above, photon frequencies were chosen to match actual experimental setup, such that their values do not exactly coincide with the excitation energies considered here.
Nonetheless, our results reproduce the same trends in terms of relative Raman shifts and resonance intensities upon increasing energies of impinging photons.
In the specific case of AB bilayer graphene, the study in Ref.~\cite{herziger} includes contributions by phonon modes along the $\Gamma$-M path which are not accounted for in the present work.

\section{\label{sec:conclusion}Conclusions}

In summary, we have provided a state-of-the-art fully \textit{ab initio} description~\cite{proceedings,ambr+97zpb} of first- and second-order Raman spectra of graphene, bilayer graphene, and graphite, in both AA and AB stacking.
In accordance with previous work~\cite{malard2009raman,ferrari_nano} we have found that the first-order G-band gives rise to the same signal in all structures, regardless of the stacking, while the second-order 2D-band is very sensitive to the specific layer sequence~\cite{venezuela2011,ferrari_layers,disorder,herziger,popov_conference,double-resonant-abc}.
Going beyond that, we have compared second-order Raman spectra of representative graphene-based materials excited by the same set of photon frequencies.
Our results indicate a significant dependence of the Raman resonance intensities on the applied excitation energy.
Peaks are consistently shifted to higher frequencies upon higher excitation energies irrespective of the considered stacking.
Our findings show that specific graphene-based materials behave differently due to the respective weight of the contributing phonon modes.
In particular, in graphene and AA graphite the 2D-band is formed by a single peak, while in bilayer graphene and AB graphite it consists of three sub-peaks given by the contributions of different phonon processes related to the iTO bands.
The second-order Raman spectrum of AA graphene, which, to the best of our knowledge, has not been computed before, is dominated by one resonance, showing evident similarities with its periodic counterpart (AA graphite) and with monolayer graphene. 
The first-principles methodology for first- and second-order Raman scattering presented and adopted in this work is applicable to any material, as it can naturally incorporate a many-body treatment of electronic excitations.
As such, it is suitable for identifying and interpreting the Raman fingerprints of emerging materials, including low-dimensional semiconductors with pronounced excitonic effects.

Input and output data of all first- and second-order Raman scattering calculations can be downloaded free of charge from the NOMAD repository at the following links: \url{http://dx.doi.org/10.17172/NOMAD/2018.06.05-2} and \url{http://dx.doi.org/10.17172/NOMAD/2018.06.03-1}.

\section*{Acknowledgements}
Work partly funded by the German Research Foundation (DFG) through the Collaborative Research Centers 658 and 951.
A.$\,$H.~acknowledges financial support from the Elsa-Neumann Foundation.
Computing time granted by the North-German Supercomputing Alliance (HLRN) on the HLRN-III system.

\section{Appendix A: Calculation of the dielectric function}
\label{app:epsilon}

For the evaluation of Eqs.\:(\ref{eq:29a}) and (\ref{eq:29b}), the macroscopic dielectric function without momentum transfer needs to be calculated in different geometries.
The macroscopic dielectric function is connected to the head element of the symmetrized reducible polarizability matrix $\boldsymbol \chi$ in the basis of reciprocal lattice vectors $\mathbf{G}$ by~\cite{kohn_response}
\begin{equation}
  \label{eq:68b}
  \varepsilon(\omega)=\frac{1}{
    1+\chi_{00}(\omega)}~.
\end{equation}
In the RPA~\cite{rpa},
the symmetrized reducible polarizability fulfills the following Dyson equation
\begin{equation}
  \label{eq:100c}
\chi^\text{RPA}_{\mathbf{G}\mathbf{G}'}(\omega)= {P}_{\mathbf{G}\mathbf{G}'\!}^0(0,\omega)+\sum_{\mathbf{G}''}   {P}_{\mathbf{G}\mathbf{G}''\!}^0(0,\omega) \, \chi_{\mathbf{G}''\mathbf{G}'\!}^{\text {RPA}}(\omega)\, ,
\end{equation}
with the symmetrized irreducible IP polarizability
\begin{align}\nonumber
  \label{eq:64}
  P^0_{\mathbf{G}\mathbf{G}'}(\mathbf{q},\omega)=&\frac{1}{N_\mathbf{k}\,V_E}\sum_{nm\mathbf{k}}
  \frac{f_{n\mathbf{k}}-f_{m\mathbf{k}+\mathbf{q}}}
                                                      {\epsilon_{n\mathbf{k}}-\epsilon_{m\mathbf{k}+\mathbf{q}}+\omega+i\eta}\\
  &\times
  M_{nm\mathbf{k}}(\mathbf{q},\mathbf{G}) \, M^*_{nm\mathbf{k}}(\mathbf{q},\mathbf{G}')~.
\end{align}
Here, $f_{n\mathbf{k}}$ is the occupation number of the single-particle Bloch state $\ket{n \mathbf{k}}$ with energy $\epsilon_{n\mathbf{k}}$. The matrix elements in Eq.\:\eqref{eq:64} are given by
\begin{align}
  \label{eq:66}
  M_{nm\mathbf{k}}(\mathbf{q},\mathbf{G})& =  \frac{\sqrt{4\pi}}{|\mathbf{q}+\mathbf{G}|}
  \mel**{n\,\mathbf{k}}{e^{\mathrm{i}(\mathbf{q}+\mathbf{G})\cdot\mathbf{r}}}{m\,\mathbf{k}+\mathbf{q}}. 
\end{align}
In the long-wavelength limit (at small \textbf{q}), using $\mathbf k \! \cdot \! \mathbf p$ perturbation theory~\cite{draxl_epsilon_ip}, the matrix elements are given by
\begin{align}
  \label{eq:66b}
  M_{nm\mathbf{k}}(\mathbf{q},0)& = 
                         \sqrt{4\pi}\; \mathbf{q} \cdot \frac{\mel**{n\,\mathbf{k}}{\mathbf{p}}{m\,\mathbf{k}}}
                         {\epsilon_{n\mathbf{k}}-\epsilon_{m\mathbf{k}}} \:.
\end{align}

\section{Appendix B: Computational details}
\label{app:comp}
All calculations are performed using the full-potential all-electron code \texttt{exciting}~\cite{exciting}, which adopts linearized augmented planewaves (LAPW) and local orbitals as basis functions. 
The local-density approximation for the exchange-correlation potential is adopted within the Perdew-Wang parameterization~\cite{LDA}.
For carbon the muffin-tin radius $R_\text{MT}\!=\!\SI{0.635}{\angstrom}$ is used.
The equilibrium geometry with in-plane C-C distance of \SI{1.412}{\angstrom} and inter-layer distance of \SI{3.324}{\angstrom} (\SI{3.632}{\angstrom}) for AB (AA) stacking is found in a lattice relaxation  with a LAPW-cutoff defined by the dimensionless parameter $R_\text{MT}G_\text{max}$=9 and a $\mathbf{k}$-point grid for sampling the BZ of 25$\times$25 for graphene and bilayer graphene, 25$\times$25$\times$16 for AA graphite, and 25$\times$25$\times$9 for AB graphite.
For graphene and bilayer graphene, a distance of \SI{7.34}{\angstrom} between the crystals and their replicas in the out-of-plane direction is used.
Phonon frequencies and eigenvectors are calculated with $R_\text{MT}G_\text{max}$=6 and a 20$\times$20 $\mathbf{k}$-grid for finite momentum phonons. For the Raman-active $\Gamma$-point phonons, the $\mathbf{k}$-grid is increased up to 80$\times$80 for AA bilayer graphene to ensure convergence within \SI{1}{cm^{-1}}. 
For calculating derivatives, a very smooth dielectric function is necessary, which requires an much denser in-plane $\mathbf{k}$-mesh than the one used for ground-state calculations. Here, a grid of 150$\times$150 $\mathbf{k}$-points, randomly shifted from the origin, and an electronic broadening of \SI{0.14}{eV} is used. Derivatives are obtained by calculating the dielectric tensor in nine geometries with a sum of absolute displacements of up to \SI{0.09}{bohr} per unit cell and using a fourth-order polynomial fit.
For second-order Raman calculations, the second derivatives of the dielectric function are calculated at five points along the $\Gamma$-K line: $\SI{1/2}{K}$, $\SI{3/5}{K}$, $\SI{2/3}{K}$, $\SI{3/4}{K}$, and $\SI{6/7}{K}$. Between the points an interpolation of the Gaussian fitting parameters is used to calculate the second-order Raman spectra. 

%


\end{document}